\documentclass[useAMS,usenatbib]{mn2e}
\usepackage{times}
\usepackage{rotating}
\usepackage{graphicx}
\usepackage{amsmath}
\long\def\symbolfootnote[#1]#2{\begingroup%
\def\thefootnote{\fnsymbol{footnote}}\footnote[#1]{#2}\endgroup}
\newcommand{\gae}{\lower 2pt \hbox{$\, \buildrel {\scriptstyle >}\over {\scriptstyle
\sim}\,$}} 
\newcommand{\lae}{\lower 2pt \hbox{$\, \buildrel {\scriptstyle <}\over {\scriptstyle
\sim}\,$}}

\begin{document}

\title[Afterglow from relativistic shock breakout]{The afterglow of a
relativistic shock breakout and low-luminosity GRBs}  

\author[Barniol Duran et al.]{R. Barniol Duran$^1$\thanks
{E-mail: rbarniol@phys.huji.ac.il, udini@wise.tau.ac.il, tsvi@phys.huji.ac.il,
  sari@phys.huji.ac.il},
E. Nakar$^2$\footnotemark[1], T. Piran$^1$\footnotemark[1] and R. Sari$^1$\footnotemark[1] \\
$^{1}$Racah Institute of Physics, Edmund J. Safra Campus, Hebrew University of Jerusalem, Jerusalem 91904, Israel \\
$^{2}$The Raymond and Berverly Sackler School of Physics and Astronomy, Tel Aviv University, 69978 Tel Aviv, Israel} 

\date{Accepted; Received; in original form 2014 July}

\pubyear{2015}

\maketitle

\begin{abstract}
The  prompt emission  of low  luminosity  gamma-ray bursts  ({\it ll}GRBs) 
indicates that these events originate  from  a  relativistic shock  breakout.
In this case, we can estimate, based on the properties of the prompt emission, the 
energy distribution of the ejecta.  We develop  a general formalism to
estimate the afterglow produced by synchrotron emission from the forward shock
resulting from the interaction of this ejecta with the circum-burst matter. 
We assess whether this emission  can  produce the observed  radio and  X-ray
afterglows of the available  sample of  4 {\it ll}GRBs.  All 4  radio afterglows  
can be  explained  within this  model, providing further support for  shock
breakouts being the origin of {\it ll}GRBs.  We find that in one of the {\it ll}GRBs (GRB 031203),
the predicted X-ray emission,  using the same parameters that fit the radio,
can explain the observed one.  In another one (GRB  980425), the observed
X-rays can be  explained if we allow  for a slight modification of the
simplest model. For the last two cases (GRBs 060218 and 100316D), we find
that, as is the case for previous attempts to model these afterglows, the
simplest model that fits the radio emission underpredicts the observed X-ray
afterglows. Using general arguments, we show that the most natural location of the X-ray 
source is, like the radio source, within the ejecta-external medium interaction layer 
but that emission is due to a different population of electrons or to a different emission process.  
\end{abstract}


\section{Introduction}

Low luminosity gamma-ray bursts ({\it ll}GRBs), or sub-energetic GRBs, are a
subsample of GRBs that emit $10^{46}$ -- $10^{48}$ erg/s of gamma-rays or hard
X-rays.  This  is several orders of magnitude below the typical luminosity of
long GRBs. Additionally, these bursts are characterized by a single pulsed
smooth light curve and by a very soft spectrum. Even though {\it ll}GRBs are more
numerous than typical long GRBs \citep{coward05, cobbetal06,
  pianetal06, soderbergetal06, liangetal07, guettaanddellavalle07,
  virgilietal09, wandermanandpiran10, fanetal11}, because of their
low luminosity only a few of these {\it ll}GRBs have been discovered to date. 
These include GRB 980425 \citep{galamaetal98, kulkarnietal98, pianetal00, 
kouveliotouetal04}, GRB 031203 \citep{malesanietal04, soderbergetal04, watsonetal04}, 
GRB 060218 \citep{campanaetal06, mazzalietal06, pianetal06, soderbergetal06} and 
GRB 100316D \citep{fanetal11, starlingetal11, marguttietal13}. {\it ll}GBRs also
have a radio afterglow, which indicates a comparable energy in mildly relativistic 
ejecta \citep{kulkarnietal98, soderbergetal04, soderbergetal06, marguttietal13}.  This
energy is only a small fraction of the total explosion energy in the supernovae 
($10^{52}$ -- $10^{53}$ erg) that follow {\it ll}GRBs.  

Several authors have suggested that 
{\it ll}GRBs are produced by shock breakouts \citep{kulkarnietal98, matznerandmckee99, tanetal01, campanaetal06, waxmanetal07, wangetal07, katzetal10}. In particular, \citet{brombergetal11}
have shown that {\it ll}GRBs' jets are not powerful enough to punch a hole in a
stellar envelope and as such they cannot be produced via the common Collapsar
model for long GRBs. At the same time, by studying the emission 
from a relativistic shock breakout, \citet{nakarandsari12} have provided strong evidence 
that the ``prompt'' emission in {\it ll}GRBs is produced by this mechanism.

The breakout of a shock that travels outward through a stellar envelope produces
the first observable light from a stellar explosion.  As the shock crosses 
the stellar envelope, and as the envelope's density decreases outwards, the
shock accelerates and it breaks out at a larger velocity than it had in 
the stellar interior, carrying, however,  only a small fraction of the explosion energy.
Before the breakout, the shock is radiation-dominated.  At the time
of the breakout, all photons in the transition layer escape and produce a short
and bright flash. Newtonian shock breakout has been explored before by many authors 
\citep{colgate74, falk78, kleinandchevalier78, imshenniketal81, ensmanandburrows92, 
matznerandmckee99, katzetal10, katzetal12, nakarandsari10}.  For relatively slow
shocks, the radiation in the shock breakout layer is in thermal equilibrium and 
it produces a signal that peaks in the UV.  For faster radiation-mediated shocks, 
the radiation is not in thermal equilibrium \citep{weaver76},  and  it produces a signal that 
peaks in X-rays \citep{katzetal10, nakarandsari10}. The signal from a
relativistic shock breakout has only been recently 
studied by \citet{nakarandsari12}, and it has been shown to produce a flash of soft gamma-rays.
In particular, they found a ``relativistic breakout closure relation'' between the
breakout energy, temperature and duration. Remarkably, the observed 4 {\it
  ll}GRBs satisfy this relationship, strengthening their interpretation in the
context of the relativistic shock breakout model. 

Following the relativistic shock breakout the envelope
expands and accelerates in a way such that outer parts of the envelope are
faster and less energetic, and inner parts are slower and more energetic. This
ejecta interacts with the circumstellar medium and drives an external shock, which
accelerates particles and generates synchrotron emission. As  slower and more
energetic material
catches up with the decelerating ejecta it re-energizes the blast wave. 
In this paper, we estimate the expected synchrotron  afterglow signal from
the forward shock  that arises from the interaction of these ejecta with the 
circum-burst material.  We then compare the expected signal with the available
multiwavelength  {\it ll}GRB afterglow data.  This can
serve as an additional test of the relativistic shock breakout model for the
``prompt'' signal in {\it ll}GRBs. 

The expected afterglow from a relativistic shock breakout is similar to the 
afterglow that arises from the long-lasting blast wave in regular GRBs
\citep{sarietal98}.  
The main difference arises from the fact that within the shock breakout  
the blast wave is continuously re-energized by slower ejecta.
The relativistic shock breakout model provides the 
distribution of the kinetic energy of the ejecta as a function of its
velocity (or its Lorentz factor). Using this distribution and 
the  ``refreshed shock model'' considered before in the context of GRBs 
(e.g., \citealp{panaitescuetal98, reesandmeszaros98, cohenandpiran99, KumarPiran00,
  sariandmeszaros00}), we estimate the resulting dynamics of the blast wave and
its emission. 

We note that the relativistic shock breakout scenario and the interaction of
the ejecta with the external medium has been considered before by
\citet{nakamura06}, and recently considered in the context of
the breakout from neutron star mergers \citep{kyutokuetal14}.  
However, in this work we consider it in the context of {\it ll}GRBs, for which we have evidence suggesting 
that this mechanism  takes place. Moreover, for {\it ll}GRBs we have available multiwavelength data 
that allow us to put this theory to the test.  We also consider not only the relativistic phase, 
but also the non-relativistic phase, which is important in some {\it ll}GRBs, where the 
shock breakout is mildly relativistic. 

The paper is organized as follows.  In Section \ref{dynamics}, we consider the dynamics of the 
relativistic shock breakout scenario.  We calculate, in Section \ref{afterglow_signal},  the light curves
and spectra expected as the ejected material following the breakout interacts with the external medium, 
both in the relativistic and the non-relativistic phases. In Section
\ref{application}, we calculate the expected afterglow signal from {\it ll}GRBs using
the general formalism presented in Section \ref{afterglow_signal} and compare
it with the multiwavelength afterglow observations of {\it ll}GRBs. We discuss
each {\it ll}GRB in Section \ref{application}. We examine the constraints on the X-ray emitting region in 
section \ref{Xray_constraints_section}. We discuss the implications of these
results and our conclusions in Section \ref{final_section}. 

\section{Shock Breakout Dynamics} \label{dynamics}

We begin summarizing the dynamics of the shock propagation within a stellar envelope before
and after the breakout \citep[e.g.,][]{johnson&mckee71,
tanetal01,nakayama05,pan&sari06}.  Before the breakout, a relativistic
shock accelerates as
\begin{equation} \label{acceleration_LF}
\gamma_i \propto \rho^{-\mu},
\end{equation}
where $\mu = \sqrt{3} -3/2 \approx 0.23$. Here, $\rho$ is the pre-shock
density and $\gamma$ is the Lorentz factor (the subscript ``i'' stands for ``initial'').  
We treat the whole expanding envelope as a series of successive
shells.  Following the breakout, these shells expand and accelerate to a final
Lorentz factor given by
\begin{equation} \label{gamma_f}
\gamma_f = \gamma_i^{1+\sqrt{3}} \propto \rho^{-0.63}.
\end{equation}
The density near the stellar edge is given by $\rho \propto z^n$, 
where $z=R_*-r$, with $R_*$ being the stellar radius and $r$ the distance from
the star center, and $n$ the polytropic index.  Typically, for the stars in which shocks will be
relativistic, $n \approx 3$.  

The mass of each of the shells at the edge is approximately given by
$m= 4 \pi R_*^2 \rho z \propto z^{n+1} \propto \rho^{1+1/n}$. The energy of
each shell after it has reached $\gamma_f$ is $E = m c^2 \gamma_f$, where $c$
is the speed of light. Using eq. (\ref{gamma_f}), we can express $m$ as a 
function of $\gamma_f$ and find $E$ as a function of $\gamma_f$ 
as follows
\begin{equation} \label{E} 
E \propto \gamma_f^{-(0.58+1.58/n)} \propto \gamma_f^{-1.1},
\end{equation}
where we have used $n=3$ in the last expression. This expression is consistent
with, e.g., \citet{tanetal01} and \citet{nakayama05}.  

\section{The Afterglow signal} \label{afterglow_signal}

As each one of the shells interacts with the circumstellar medium, they
will drive a forward and a reverse shock. Particles will be accelerated in
these shocks and radiation will be produced.  In the scenario considered
above, shells will continuously inject energy to the blast wave, since slower shells contain
more energy than faster ones and they catch up with the decelerating ejecta 
(e.g., \citealp{panaitescuetal98, reesandmeszaros98, cohenandpiran99, 
KumarPiran00, sariandmeszaros00}). 
The energy injection, in this scenario is typically characterized as 
$E \propto \gamma_f^{-s+1}$. The results of the last section show that, for a 
relativistic breakout from a stellar envelope 
\begin{equation} \label{s}
s=1.58(1+1/n)=2.1,
\end{equation}
where we have used $n=3$ in the last expression.

\subsection{Energetics} \label{Energetics}

At the time of breakout, most of the energy is emitted by a shell with
pre-explosion optical depth of order unity.  We  denote this shell as
the {\it breakout shell}, and define its final Lorentz factor as
$\gamma_{f,0}$ and its final energy as $E_{f,0}$. Subsequent shells
will be more energetic but slower than the breakout shell. The final Lorentz
factor of the breakout shell is crucial in determining the total bolometric 
energy radiated from the breakout and the duration/temperature of the breakout
emission. It allows for the derivation of a ``relativistic breakout closure relation''
\citep{nakarandsari12}. The optical depth of the
breakout shell, $\tau_0 \approx \kappa_T m_0/(4 \pi R_*^2)=1$, with $\kappa_T
\approx 0.2$ cm$^2$ g$^{-1}$, implies that the mass of the breakout shell 
is \citep{nakarandsari12}:
\begin{equation}
m_0 = 4\times10^{-9} M_{sun} R_5^2,
\end{equation}
where $R_5$ is the stellar radius in units of $5 R_{\odot}$. Using   
$E_{f,0}=m_0 c^2 \gamma_{f,0}$ we find
\begin{equation} \label{energy_norm}
E_{f,0} =(7.2 \times 10^{45} \, {\rm erg}) R_5^2 \gamma_{f,0}.
\end{equation}
Therefore, we can normalize the proportionality relation  
eq. (\ref{E}) using $E_{f,0}$ and $\gamma_{f,0}$ as 
\begin{equation} \label{E_vs_gamma}
E = E_{f,0} \left(\frac{\gamma_f}{\gamma_{f,0}}\right)^{-s+1}.
\end{equation}
In summary, by identifying the emission from the breakout and using the
relativistic breakout relation, we can determine $\gamma_{f,0}$ and the
breakout stellar radius.  With these two quantities, we can determine
$E_{f,0}$ using eq. (\ref{energy_norm}) and  determine how much energy 
is injected to the blast wave using eq. (\ref{E_vs_gamma}). 

\subsection{The relativistic phase} \label{Relativistic_phase}

The breakout shell, with energy $E_{f,0}$ and Lorentz factor $\gamma_{f,0}$,
will decelerate in a time (see, e.g., \citealp{panaitescuandkumar00}):
\begin{equation} \label{t_dec0} 
t_{dec,0} = \begin{cases} 0.7 \, {\rm d} 
\left(\frac{E_{f,0}}{10^{50} \, {\rm erg}} \frac{1 \, {\rm cm^{-3}}}{n}\right)^{\frac{1}{3}} 
\left(\frac{\gamma_{f,0}}{5}\right)^{-\frac{8}{3}} (1+z) & \mbox{ISM,} \\
0.01 \, {\rm d} 
\left(\frac{E_{f,0}}{10^{50} \, {\rm erg}}\right) A_{*}^{-1} \left(\frac{\gamma_{f,0}}{5}\right)^{-4} (1+z)
& \mbox{wind,}  
\end{cases}
\end{equation}
as it interacts with an external density given by $n \propto R^{-k}$ (not to be
confused with the polytropic index), where
$R$ is the distance from the center of the explosion.
For a constant density
medium (ISM) $k=0$, and for a wind medium $k=2$, where $A_{*}$ is the wind
parameter\footnote{The wind parameter is defined as 
$A=\dot{M}_w/ 4 \pi V_w = 5\times 10^{11} A_{*}$ g cm$^{-1}$, where 
$\dot{M}_w$ is the wind mass-loss rate and $V_w$ is its velocity.  
The reference value for $A$ was scaled using  
$\dot{M}_w=10^{-5}$ M$_{sun}$ yr$^{-1}$ and $V_w = 1000$ km s$^{-1}$.} in units of $5\times10^{11}$ g cm$^{-1}$.  
The slower/more energetic shells will catch up with the faster/less energetic
ones when the latter have decelerated to a comparable Lorentz factor.  
At this point, the Lorentz factor of the  shocked material is \citep{sariandmeszaros00}:
\begin{equation} \label{gamma_f_vs_time}
\gamma_f = \gamma_{f,0} \left(\frac{t}{t_{dec,0}}\right)^{-\frac{3-k}{7+s-2k}},
\end{equation}
and the blast wave energy increases as
\begin{equation} \label{E_vs_time}
E = E_{f,0} \left(\frac{t}{t_{dec,0}}\right)^{\frac{(3-k)(s-1)}{7+s-2k}},
\end{equation}
where time $t$ is the observer's time since breakout.  
Thus, due to  the energy injection of slower/more energetic shells, 
the blast wave LF decrease with time slower compared with the case when no
energy injection proceeds ($s=1$).  For the case considered here, $s=2.1$, the
LF decreases as $\propto t^{-0.33}$ ($\propto t^{-0.2}$) for ISM
(wind) medium. Meanwhile, the blast wave energy increases as, see
eq. (\ref{E_vs_time}), $\propto t^{0.36}$ 
($\propto t^{0.22}$) for the ISM (wind) case.

The {\it relativistic phase} of the afterglow, for the typical parameters
considered here, will last until a time, $t_{NR}$, when $\gamma_f \approx 1$
in eq. (\ref{gamma_f_vs_time}):
\begin{equation} \label{t_NR}
t_{NR} = \begin{cases} 92 \, {\rm d} 
\left(\frac{E_{f,0}}{10^{50} \, {\rm erg}} \frac{1 \, {\rm cm^{-3}}}{n}\right)^{\frac{1}{3}} 
\left(\frac{\gamma_{f,0}}{5}\right)^{0.4} (1+z) 
& \mbox{ISM,} \\
37 \, {\rm d} \left(\frac{E_{f,0}}{10^{50} \, {\rm erg}}\right) A_{*}^{-1} 
\left(\frac{\gamma_{f,0}}{5}\right)^{1.1} (1+z) & \mbox{wind}. \end{cases}
\end{equation}
Afterward, the energy injection of non-relativistic shells will be
important. 

We note that acceleration continues beyond the breakout shell,
up to the outermost shell, as long as the temperature in the shock exceeds
$\sim 50$ keV, which is required for the creation of a significant amount of
pairs (see \citealp{nakarandsari12} for details).  That is, although the breakout shell carries most of
the energy during the breakout event, it is preceded by shells with lower energy
-- as prescribed by eq. (\ref{E_vs_gamma}) -- but that move faster.  A conservative lower
limit of the {\it final} Lorentz factor of the
outermost shell is given by \citep{nakarandsari12}
\begin{equation} \label{gamma_max}
\gamma_{f,max} \approx 4.3 \gamma_{f,0}^{0.65}.
\end{equation}
This faster material will decelerate earlier than $t_{dec,0}$, 
since the deceleration time depends much stronger on the
Lorentz factor  than on the energy (see eq. \ref{t_dec0}).
These estimates are limited to $\gamma_{f,0} \sim 30$, which 
corresponds to $\gamma_{f,max} \sim 40$ (see \citealp{nakarandsari12}).

\subsection{The non-relativistic phase} \label{NR phase}

After a time $t_{NR}$ the blast wave velocity becomes
non-relativistic.  Slower shells continue to inject
energy to the blast wave, similarly as in Section \ref{dynamics}, but the shock
velocity -- instead of the Lorentz factor -- increases as $\propto
\rho^{-\mu}$, see eq. (\ref{acceleration_LF}), where $\mu \approx 0.19$ \citep{sakurai60}.
Following a similar procedure as in Section \ref{dynamics}, it can be shown
that the energy increases as a steep function of
the velocity $\beta c$ \citep{tanetal01}
\begin{equation} \label{s'}
E \propto (\beta \Gamma)^{-(3.35+5.35/n)},
\end{equation}
which for $n=3$, as considered above, yields $E \propto (\beta
\Gamma)^{-s'+1} = (\beta \Gamma)^{-5.2}$, where $s'$ in the
non-relativistic phase satisfies $s'= 4.35+5.35/n \approx 6.2$.  
To take into account energy injection  to the external shock  we follow the same procedure as in
\citet{sariandmeszaros00}.  The total energy in the blast wave is 
\begin{equation}
E \sim M(R) (\beta c)^2,
\end{equation}
where $M(R)$ is the mass of the collected external medium 
up to radius $R$.
The velocity decreases and the energy increases with time as:
\begin{equation}
\beta \propto t^{-\frac{3-k}{4-k+s'}},
\end{equation}
and
\begin{equation} \label{E_non_rel}
E \propto t^{\frac{(3-k)(s'-1)}{4-k+s'}}.
\end{equation}
For the case considered here, $s'=6.2$, the velocity decreases slowly  as 
$\propto t^{-0.29}$ ($\propto t^{-0.12}$), while the blast wave energy
increases as $\propto t^{1.53}$ ($\propto t^{0.63}$) for the ISM (wind) case.

We note that the energy injection in the non-relativistic phase will proceed
until a characteristic velocity of the order of $\sim \sqrt{E/M}$ is reached, where $E$ and $M$ are 
the total explosion energy and the total ejected mass, respectively.

\subsection{The spectra and light curves} \label{Spectra}

We turn now to estimate the spectra and light curves expected from these refreshed shocks.

\subsubsection{The coasting phase afterglow} \label{coasting}

Initially, until  the deceleration of the 
fastest ``shell'', see eq. (\ref{gamma_max}), this shell coasts at a constant 
velocity and  the afterglow is
dominated by its  emission. This phase is called the
``coasting phase''.  Synchrotron emission from this phase has been obtained before
(see, e.g., \citealp{sari97} and recently, \citealp{nakar&piran11,shen&matzner12}). 
Here, we present the light curves for $\nu_a < \nu_m < \nu_c$ (other cases can
be found easily). For the ISM [wind] case, the light curves are

\begin{equation} \label{F_nu_coasting} 
F_{\nu} \propto \begin{cases} t^2 [t^2] \nu^2 & \mbox{$\nu < \nu_m$} \\
t^2 [t^{5/2}] \nu^{5/2} & \mbox{$\nu_m < \nu < \nu_a$} \\
t^3 [t^{-\frac{(p-1)}{2}}] \nu^{-\frac{(p-1)}{2}} & \mbox{$\nu_a < \nu < \nu_c$} \\
t^2 [t^{-\frac{(p-2)}{2}}] \nu^{-\frac{p}{2}} & \mbox{$\nu_c < \nu$.} 
\end{cases}
\end{equation}
Since the velocity during this phase is constant in time, these light curves are
valid both for a relativistic and a non-relativistic shell.
{The light curves and spectra presented in the next two subsections are valid {\it after} the
deceleration of the fastest ``shell'', when the blast wave energy starts to 
be refreshed by slower moving ejecta.}

\subsubsection{The relativistic afterglow} 

Using the formalism presented in \citet{sariandmeszaros00} one can calculate
the spectra and light curves of the forward shock synchrotron
emission when energy injection occurs as
described above (that is, for a given set of parameters: $E_{f,0}$,
$\gamma_{f,0}$ and $s$). \citet{sariandmeszaros00} do provide a rough estimate
of the synchrotron emission from the reverse shock, however, in order to
calculate this emission more accurately, one needs to consider the exact numerical
calculation of the strength of the reverse shock presented in \citet{nakamura06} for the relativistic
case and in \citet{chevalier82} for the non-relativistic (Newtonian) case. 
The density behind the reverse shock is larger than that behind the forward
shock, however, the reverse shock will peak at a much lower frequency than
that of the forward shock. For this reason, we ignore the emission from the
reverse shock and focus on the forward shock emission. In Section
\ref{application}, we will show that the forward shock model fits the radio
data of {\it ll}GRBs, and thus, the accompanying reverse shock synchrotron emission would be at
even lower frequencies.

For a given set of afterglow parameters: density
($n \propto R^{-k}$), power-law index of energy distribution of electrons ($p$) and 
microphysical parameters ($\epsilon_e$ and $\epsilon_B$), we can use 
\citet{sariandmeszaros00} to predict the expected afterglow emission at any
wavelength and time. We follow a more precise procedure, which is to 
calculate the energy of the  blast wave at any given time and, with it, calculate the forward shock
emission using \citet{granotandsari02}, since this last work presents more precise
numerical coefficients in the forward shock synchrotron emission calculation.
The predicted spectrum will be given by
the synchrotron spectrum of a power-law distribution of electrons. It is
characterized by three frequencies: the self-absorption frequency, $\nu_a$, the minimum
frequency, $\nu_m$, and the cooling frequency, $\nu_c$.  

To sketch the calculation, we provide the scalings of the characteristic
frequencies as a function of time and energy for a general density
stratification (for compactness), although we focus on the interstellar medium (ISM) and wind
cases. The energy at a given time,
as a function of $s$ and $k$, is given by eq. (\ref{E_vs_time}).  For the case of 
$\nu_a < \nu_m < \nu_c$ the peak flux is $F_m$ at $\nu_m$. The specific flux at each of the
synchrotron segments is $\propto \nu^2$, $\nu^{1/3}$, $\nu^{-(p-1)/2}$,
$\nu^{-p/2}$. We have (see, e.g., \citealp{granotandsari02, leventisetal12})
\begin{eqnarray}
\nu_a &\propto& t^{-\frac{3k}{5(4-k)}}E^{\frac{4-4k}{5(4-k)}} \nonumber \\
\nu_m &\propto& t^{-\frac{3}{2}} E^{\frac{1}{2}}  \\
\nu_c &\propto& (t E)^{-\frac{4-3k}{2(4-k)}} \nonumber \\
F_m &\propto& t^{-\frac{k}{2(4-k)}}E^{\frac{8-3k}{2(4-k)}}. \nonumber 
\end{eqnarray}
For the case of $\nu_m < \nu_a < \nu_c$, the peak flux is at $\nu_a$
(see below) and given by $F_m (\nu_a/\nu_m)^{-(p-1)/2}$. The specific flux at each of the
synchrotron segments is $\propto \nu^2$, $\nu^{5/2}$, $\nu^{-(p-1)/2}$,
$\nu^{-p/2}$. Both $\nu_m$ and $\nu_c$ behave as above. However, $\nu_a$ is given by
\begin{equation}
\nu_a \propto t^{\frac{3kp-2k-12p-8}{2(4+p)(4-k)}}E^{\frac{8+4p-kp-6k}{2(4+p)(4-k)}}.
\end{equation}
Light curves at a fixed frequency are obtained by constructing the synchrotron
spectrum with the given characteristic frequencies and letting them evolve
with time \citep{sarietal98}. For simplicity, we do not consider the 
smoothing between different power-law segments in the synchrotron spectrum
(see, e.g., \citealp{granotandsari02,leventisetal12}).

We now provide the ``closure relations'', which is a relation between the
temporal decay index, $\alpha$, and the spectral index, $\beta$ (not to be
confused with the velocity of the ejecta), where
the specific flux at frequency $\nu$ is defined as $f_{\nu} \propto
t^{-\alpha} \nu^{-\beta}$.  We focus on the light curves above the peak
frequency (the other regimes can be found easily with the equations above),
and find
\begin{equation} \label{closure1}
\alpha = \begin{cases} \frac{-\beta(24-7k+sk)-6+6s+k-3sk}{4k-2(7+s)} & \mbox{$\nu_m < \nu < \nu_c$}\\
\frac{-\beta(24-7k+sk)-(k-4)(1+s)}{4k-2(7+s)} & \mbox{$\max(\nu_m,\nu_c) < \nu$.} \end{cases}
\end{equation}
These expressions are consistent with \citet{sariandmeszaros00}. For example,
for $p=2.4$ and $s=2.1$, the ISM (wind) case yields decaying light curves, with the
light curve for $\max(\nu_m,\nu_c) < \nu$ being steeper (shallower) than that
of $\nu_m < \nu < \nu_c$.

\subsubsection{The non-relativistic afterglow} 

The relativistic and non-relativistic solutions should be
joined at $t_{NR}$. At $t> t_{NR}$, we use the non-relativistic expressions for the
energy injection discussed above and the expressions for the 
synchrotron characteristic frequencies presented in
\citet{leventisetal12}. We note that this is an idealized case. There will be a
smooth transition both in the dynamics of the blast wave \citep{tanetal01} and 
in the exact shape of the synchrotron spectrum \citep{leventisetal12}
as the blast wave transitions from the relativistic to the non-relativistic
phases. In general, this will make $t_{NR}$ larger than reported in eq. (\ref{t_NR}).

Similarly, as done for the relativistic phase, to sketch the calculation we
provide the scalings of the characteristic frequencies as a function of time
and energy for a general density stratification (for compactness), although
we focus on the ISM and wind cases. The energy at a given time,
as a function of $s'$ and $k$, is given
by eq. (\ref{E_non_rel}).  For the case of $\nu_a < \nu_m < \nu_c$ we have 
(see, e.g., \citealp{leventisetal12})
\begin{eqnarray}
\nu_a &\propto& t^{\frac{30-16k}{5(5-k)}}E^{-\frac{5+4k}{5(5-k)}} \nonumber \\
\nu_m &\propto& t^{\frac{4k-15}{5-k}} E^{\frac{10-k}{2(5-k)}}  \\
\nu_c &\propto& t^{\frac{2k-1}{5-k}} E^{-\frac{3(2-k)}{2(5-k)}} \nonumber \\
F_m &\propto& t^{\frac{3-2k}{5-k}}E^{\frac{8-3k}{2(5-k)}}. \nonumber 
\end{eqnarray}
As for the relativistic phase, for the case of $\nu_m < \nu_a <
\nu_c$, both $\nu_m$ and $\nu_c$ remain the same.  However, $\nu_a$ is given
by 
\begin{equation}
\nu_a \propto t^{\frac{10-8k-15p+4kp}{(4+p)(5-k)}}E^{\frac{10p-kp-6k}{2(4+p)(5-k)}}.
\end{equation}

The ``closure relations'' for the non-relativistic phase are given by
\begin{equation} \label{closure2}
\alpha = \begin{cases} \frac{\beta(30-9k+s'k)-6s'+3s'k+k}{2(4-k+s')} & \mbox{$\nu_m < \nu < \nu_c$}\\
\frac{\beta(30-9k+s'k)-4(4+s')+k(5+s')}{2(4-k+s')} & \mbox{$\max(\nu_m,\nu_c) < \nu$.} \end{cases}
\end{equation}
For example, for $p=2.4$ and $s'=6.2$, the ISM (wind) case yields increasing
(decaying) light curves, with the light curve for $\max(\nu_m,\nu_c) < \nu$ 
being shallower than that of $\nu_m < \nu < \nu_c$.

\section{\MakeLowercase{{\it ll}}GRB afterglows} \label{application}

Based on the observed properties of the prompt emission of {\it ll}GRBs,
\citet{nakarandsari12} suggested that the shock breakout in these events is (mildly) 
relativistic.  In this section we compare 
the resulting afterglows, calculated with the formalism presented above and based 
on the estimated breakout properties of \citet{nakarandsari12}, 
with the afterglow observations of these {\it ll}GRBs.  

Using the prompt properties of {\it ll}GRBs, \citet{nakarandsari12} 
determine the breakout shell final Lorentz factor, $\gamma_{f,0}$, and the breakout radius $R$ 
(see Table \ref{table1}). With these quantities, we determine $E_{f,0}$, the
initial energy injected to the blast wave, using eq. (\ref{energy_norm}).  
Subsequent shells will inject more energy as prescribed in
eq. (\ref{E_vs_time}) until $t_{NR}$, when the energy injection will proceed in the
non-relativistic phase as prescribed in eq. (\ref{E_non_rel}). 
Given that these {\it ll}GRBs are accompanied by supernovae
with estimated kinetic energies between $10^{52}$ and $10^{53}$ erg, an
estimate of this energy can serve as a good quantity
of the total energy in the explosion and the point at which energy
injection should stop. We note however that this energy is not reached until a
time which is much longer than the time of the last available observations in
the radio and X-ray bands.

As in \citet{nakarandsari12}, we divide the 4 {\it ll}GRBs in two pairs with
different properties. GRBs 980425 and 031203 both have observed prompt 
peak energies in excess of $\sim 50$ keV \citep{sazonovetal04, kanekoetal07}. 
In these two GRBs, significant amount of pairs are created in the shock, 
and ejecta are accelerated to Lorentz factor larger than $\gamma_{f,0}$, 
up to $\sim \gamma_{f,max}$, see eq. (\ref{gamma_max}) \citep{nakarandsari12}.  
GRBs 060218 and 100316D both have observed prompt peak energies marginally
consistent with $\sim 50$ keV \citep{kanekoetal07, starlingetal11}.  In
these two GRBs, significant amount of pairs are only marginally created, and
ejecta with Lorentz factor larger than $\gamma_{f,0}$ is most likely 
absent. In that case the afterglow emission is produced in the coasting
phase, as described in Section \ref{coasting}, since the deceleration of the
breakout shell takes place only after very long time. Moreover, since the late 
time radio light curves of these two GRBs decrease with time, then the ejecta 
must interact with a wind medium (otherwise all light curves would increase
with time).

\begin{table}
\begin{center}
\begin{tabular}{c|cccc}
\hline
GRB & $E_{\gamma,iso}$[erg] & $R$ [cm] & $\gamma_{f,0}$ & $E_{f,0}$ [erg] \\
\hline
980425 &  $7\times10^{47}$ & $6\times10^{12}$ & 3          & $6\times10^{48}$ \\
031203 & $4\times10^{49}$ & $2\times10^{13}$ & 5 & $1\times10^{50}$ \\
060218 & $6\times10^{49}$ & $5\times10^{13}$ & $\approx$1.3 & $2\times10^{50}$\\
100316D& $6\times10^{49}$ & $5\times10^{13}$ & $\approx$1.3 & $2\times10^{50}$\\
\hline
\end{tabular}
\end{center}
\caption{Using the values of the breakout radius, $R$, and the breakout shell final
Lorentz factor, $\gamma_{f,0}$, reported in \citet{nakarandsari12}
we find the total energy of the breakout shell, $E_{f,0}$, using
Eq. (\ref{energy_norm}). The total observed isotropic energy in 
gamma-rays/hard X-rays, $E_{\gamma,iso}$, is also presented for comparison.} 
\label{table1}
\end{table}

\subsection{Light curves}

To calculate the expected light curves, 
we use the values of $E_{f,0}$ and $\gamma_{f,0}$ from Table \ref{table1} and
use $s=2.1$ and $s'=6.2$ as found above.\footnote{ 
We have assumed $n=3$, that is, a radiative stellar envelope. For a convective
envelope $n=1.5$, and $s$ and $s'$ are determined by eq. (\ref{s}) and
following eq. (\ref{s'}). In this case the temporal decay of the light curves 
above the peak are only modified by $|\Delta \alpha| \lae 0.3$
according to eqs. (\ref{closure1}) and (\ref{closure2}).}  We fix the electron
energy power-law index to $p=2.4$ (we
will comment on this point below for each burst). We also fix the fraction of shocked energy
in electrons to $\epsilon_e=0.2$, since $\epsilon_e$ seems to be narrowly
clustered around this value in afterglow modeling studies (see figure 1 in
\citealp{santanaetal14}). For cases
in which a satisfactory fit is not reached with $\epsilon_e =0.2$, we consider
other values. We calculate both light curves, for an ISM ($k=0$) with density $n$, and for a wind medium
($k=2$) with wind coefficient $A_{*}$.  We scan the parameter space in density, 
$n \sim 10^{-3} - 10^3$ cm$^{-3}$ and $A_{*} \sim 10^{-3} - 10^3$ for the ISM
and wind case, and in $\epsilon_B \sim 10^{-6} - 10^{-1}$. We look for a set
of parameters that best fits the radio observations within a
factor of $\lae 2$.  We later check if the fit to the radio light curves also gives a good fit to the
X-ray observations.  

In the following subsections, we will discuss each one of the four {\it ll}GRBs described in
Table \ref{table1} in the context of the model described above. We do not 
discuss them in chronological order, but in an order that reflects  
our overall understanding of each burst within this model.

\subsection{GRB 031203}


\begin{figure*}
\begin{center}
\includegraphics[width=16cm, angle = 0]{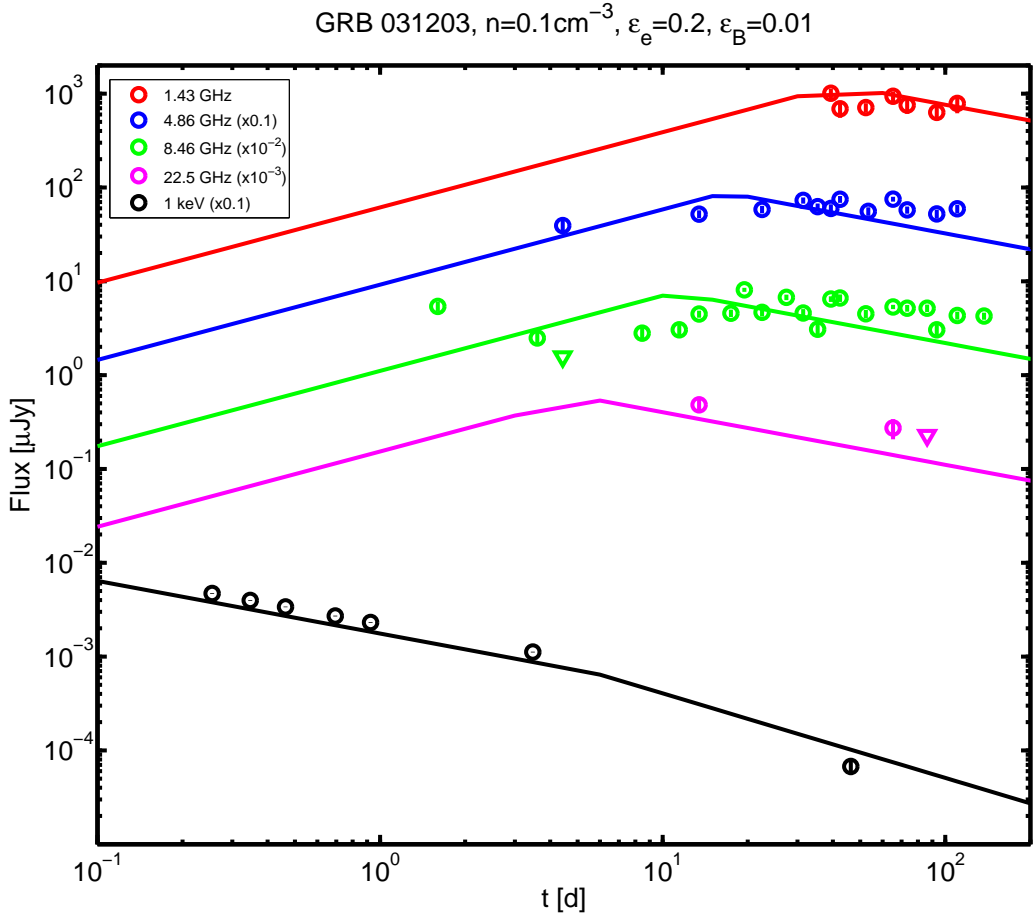}
\end{center}
\caption{The observed and predicted radio and X-ray afterglows of GRB 031203. 
  The fit shown is for $E_{f,0} = 10^{50}$ erg, $\gamma_{f,0}=5$,  $s=2.1$,
  $p=2.4$, $\epsilon_e=0.2$, $\epsilon_B=0.01$ and an ISM density with $n=0.1$cm$^{-3}$.
   Here and in the rest of the figures, circles denote observations, vertical lines within the
  circles denote flux errors and triangles are upper limits.}
\label{fig1} 
\end{figure*}


GRB 031203 was monitored extensively in the radio bands 
\citep{soderbergetal04}, and it was also followed in the X-ray band starting 6
hours after the burst \citep{watsonetal04}.  We calculate the predicted radio 
fluxes  in the ISM scenario and present them in Fig. \ref{fig1}.  
We find that the model can fit both the radio and {\it also} the X-ray
observations quite well. 

We find that the X-ray band before 6 days is above $\nu_m$ but below $\nu_c$.
It decreases as $\propto t^{-0.55 \pm 0.05}$ \citep{watsonetal04}.  The closure
relations in eq. (\ref{closure1}) yield a consistent temporal decay
with the chosen value of $p=2.4$. Within this model, the reason why the X-ray light curve steepens at $\sim
6$ days is due to the crossing of the synchrotron cooling frequency through the X-ray observing
band (the synchrotron cooling frequency decreases with time in this case). The observed
X-ray spectrum at $t < 6$ days
is $\beta_x = 0.9 \pm 0.05$ and $0.7 \pm 0.2$ at 6 hours and 3 days,
respectively \citep{watsonetal04}, roughly consistent with the predicted spectrum
$\beta_x = (p-1)/2 = 0.7$. 
The X-ray spectrum after $\sim 6$ days should steepen by 
$\Delta \beta_x =0.5$, which can be tested with the {\it Chandra} data at
$\sim 50$ days \citep{foxetal04}. 

The blast wave energy, which changes with time as energy is injected, 
is $\sim 2\times10^{50}$ erg at 10 days in this case; the time dependence of the blast wave energy
can be found in eq. (\ref{E_vs_time}).
The non-relativistic phase for this burst will commence beyond 300 days, 
which is a factor of few larger than the time of the last observations and all 
the light curves would slightly rise. We note that solutions with a slightly 
smaller density and larger $\epsilon_B$ also fit the data well and the
non-relativistic phase might be delayed for an even longer time.

The wind case is not able to fit the radio
observations, even if we allow $\epsilon_e$ to vary from the fiducial adopted
value of $\epsilon_e = 0.2$. Also, the radio light curves would be steeper than
$\propto t^{-1.1}$ for $2< p < 3$, according to eq. (\ref{closure1}), which is much 
steeper than observed (the radio light curves are quite flat at late times).
We therefore do not show a figure for this scenario.   

Interestingly, IR-optical observations starting 9 hr after the burst show a
bright IR emission with a very steep soft spectrum, $F_\nu \propto
\nu^{-2.36}$ \citep{malesanietal04}. This emission fades within a day, before
the main component of the supernova (SN) light rises a week later. The very soft spectrum
of early IR emission indicates that it has, most likely, a different origin
than that of both the radio and the X-rays. It is still consistent with the
radio and X-ray being both generated by synchrotron emission, as
our model predicts an IR flux that is fainter by a factor of $\sim 10$ than
the one observed. We do not discuss here the origin of this unique IR
emission, as it is most likely related to the cooling envelope phase of the
SN. In that case the bright yet fast fading signal may indicate on a
progenitor composed of a compact core and an extended low-mass envelope 
\citep{NakarPiro14}. 

\citet{soderbergetal04} have modeled the radio emission of this burst in the
context of the external forward shock emission (without energy
injection). They also prefer an ISM with similar density as the one we adopt here.
However,  energy injection via  the blast wave enables us  explain also the X-ray emission, which \citet{soderbergetal04} had found was a factor of $\lae 10$ smaller than observed.   

\subsection{GRB 980425}


\begin{figure*}
\begin{center}
\includegraphics[width=16cm, angle = 0]{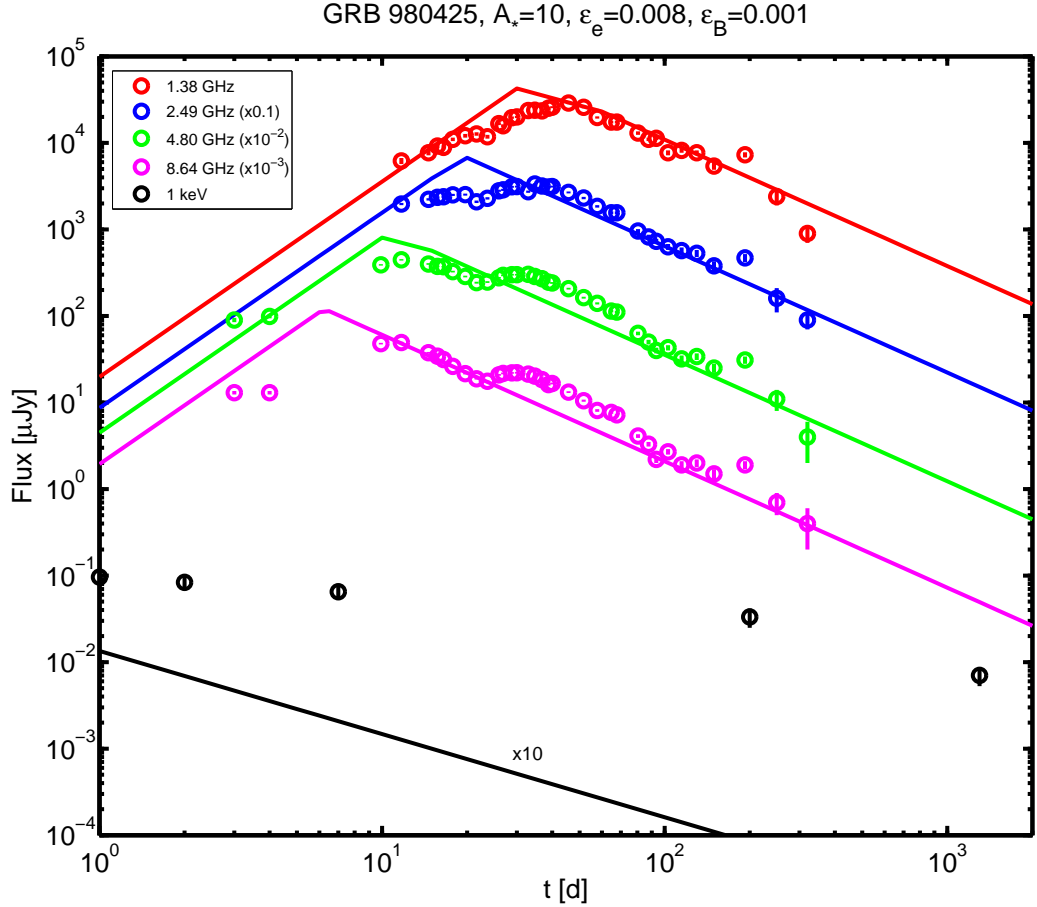}
\end{center}
\caption{
The observed and predicted radio and X-ray afterglows of GRB 980425. 
 The fit shown uses $E_{f,0} = 6 \times 10^{48}$ erg, $\gamma_{f,0}=3$, 
 $s=2.1$, $p=2.8$, $\epsilon_e = 0.008$ and $\epsilon_B=0.001$. 
  This model corresponds to the wind case with $A_* = 10$, which 
  can fit the overall trend of the radio observations (albeit with a small value of 
  $\epsilon_e$), but the X-ray prediction lies below the observations (we
  have multiplied the X-ray expectations of the model by a factor of
  10). Using the same parameters, but allowing for a smaller value of $p$ to
  model the X-rays ($p=2.1$) yields an X-ray flux consistent with observations.}
\label{fig2} 
\end{figure*}


GRB 980425 was monitored extensively in the radio bands 
\citep{kulkarnietal98}, and its X-ray data showed an unusually flat behavior for
more than 100 days \citep{pianetal00, kouveliotouetal04}. 
We use the scenario described above for the parameters of this {\it ll}GRB to model the
radio light curves. The initial blast wave energy of this {\it ll}GRB is the smallest
of all {\it ll}GRBs, therefore, we need to consider the non-relativistic
phase since $t_{NR}$ falls within the available radio observations for 
$n \gae 0.01$ cm$^{-3}$ ($A_{*} \gae 0.01$) in the ISM (wind) case, see
eq. (\ref{t_NR}). 

We consider first the ISM case.  The light curves above the minimum frequency 
in the non-relativistic phase increase with time, see eq. (\ref{closure2}), 
which is inconsistent with the time-decreasing trend of the radio data at late
times. Therefore, we consider the wind medium, for which the solution lies entirely in the
non-relativistic phase. Using a value of $\epsilon_e = 0.2$ 
as discussed above, we find that the wind medium is not able to fit the
radio observations.  We lower $\epsilon_e$ to a smaller value 
($\epsilon_e \sim 0.01$) and find that the predicted light curves give an
overall good fit to the radio observations (however, the radio observations at
$< 10$ days and the slight increase in flux at $\sim 30$ days are not well
fit). We note that the radio light curves beyond $\sim 30$ days are above $\nu_a$
and $\nu_m$, but still below $\nu_c$.  Observations indicate that they 
decrease roughly as $\propto t^{-1.6}$ \citep{liandchevalier99}.  Using the
closure relations in eq. (\ref{closure2}), we find that a value
of $p \approx 2.8$ fits the decay of the radio observations, and we
present the light curves in Fig. \ref{fig2} for this case.  
The predicted X-ray emission (which lies above $\nu_c$) is smaller than the
observed value by at least a factor of a few tens. The blast wave energy at 10 days in 
this case is $\sim 3\times10^{50}$ erg; the time dependence of the blast wave energy
can be found in eq. (\ref{E_non_rel}).

Interestingly, in the wind case and using the same parameters as above, but
changing the value of $p$ 
to $p\approx2.1$, we find an X-ray flux consistent with the observations, both in 
normalization and temporal decay.  The X-ray light curve until $\sim 200$ days 
decays as $\propto t^{-0.16 \pm 0.04}$ \citep{pianetal00}, but including the flux 
at $\sim 1000$ days, the overall light curve decays as $\sim t^{-0.33}$.  This is 
consistent with the predicted temporal decay $\propto t^{-0.4}$, 
according to eq. (\ref{closure2}).
The observed X-ray spectrum for $t \lae 200$ days is $\beta = 1.0 \pm 0.3$ 
\citep{pianetal00}, consistent with the predicted value of $\beta = p/2\approx 1.1$.
The predicted radio light curve with $p=2.1$ would be too shallow ($\propto t^{-0.9}$), but would 
qualitatively fit the radio observations within a factor of $\lae 2$ until $\sim 100$ 
days, and the discrepancy would increase up to a factor of $\lae 10$ at $\sim
300$ days. These results suggest that both the radio and the X-rays can be
generated by synchrotron emission from a single population of electrons in the
forward shock, if the electron spectrum is concave. Namely, if the electron
spectrum is steep (soft) at low energies and shallow (hard) at high
energies. Such spectrum is predicted by some {\it Fermi} acceleration models in
Newtonian shocks, due to the nonlinear feedback of the accelerated particles
on the shock structure \citep[e.g.,][]{Bell87}. 

\citet{liandchevalier99} have modeled the radio data of GRB 980425 also in the
wind case and they present several solutions: 
One with $\epsilon_e=\epsilon_B=0.1$ and $A_{*}\approx0.04$,
another with $\epsilon_e \sim 1$, $\epsilon_B=10^{-6}$ and
$A_{*}\approx6$, and they also find acceptable models with
$\epsilon_e \sim 0.01$. Our solution for this GRB has $\epsilon_e = 0.01$,
however, the two models are inherently different.  \citet{liandchevalier99} 
allow for the energy of the blast wave to increase {\it instantaneously} by a 
factor of $\sim 2.5$ to explain the increase by a factor of $\sim 2$ in the
radio light curves at $\sim 30$ days. However, within out model energy 
is {\it continuously} injected to the blast wave and 
we ignore this slight increase
in the radio light curves at  $\sim 30$ days.

\subsection{GRB 060218}


\begin{figure*}
\begin{center}
\includegraphics[width=16cm, angle = 0]{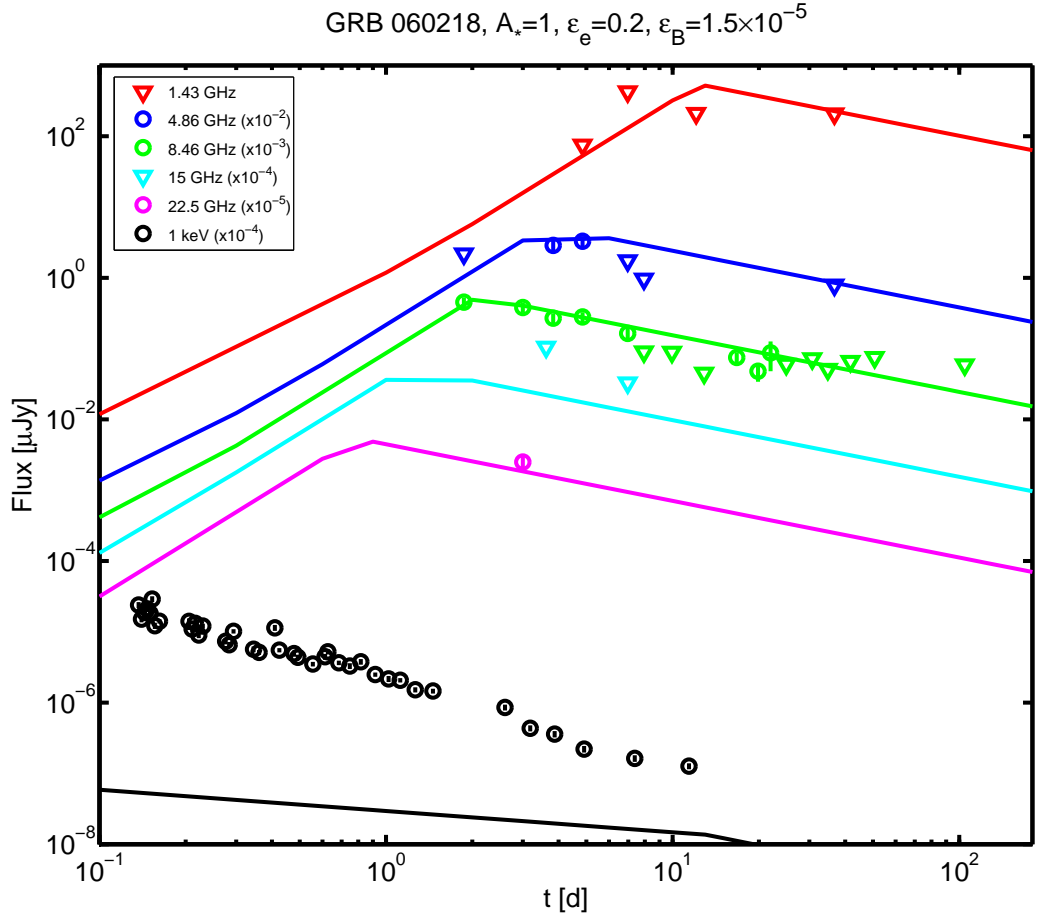}
\end{center}
\caption{The observed and predicted radio and X-ray afterglows of GRB 060218. 
  The fit shown uses $E_{f,0} = 2 \times 10^{50}$ erg, $\gamma_{f,0}=1.3$,
  $p=2.6$, $\epsilon_e=0.2$ and $\epsilon_B=1.5 \times 10^{-5}$.  This model
  corresponds to the synchrotron emission
  from the breakout shell during the coasting phase for a wind with $A_*=1$, 
  which can qualitatively fit the radio observations. 
  However, the model underpredicts the X-ray emission (and, even if the X-ray
  would not be underpredicted, the model cannot explain the steep observed X-ray spectrum).}
\label{fig3} 
\end{figure*}


The Lorentz factor of the breakout shell for this burst is close to unity and
creation of pairs in the shock is marginal. Therefore, it is likely that no
shells are accelerated to velocities faster than  the breakout shell,
in which case the forward shock is coasting at a constant velocity for a long
time until the breakout shell  decelerates. In this case only a
wind medium can fit the observed decaying radio afterglow and this is the
model we consider here. 
The blast wave moves with a constant Lorentz factor of $\sim 1.3$
until it decelerates at $\sim 200$ days. During the coasting phase 
the blast wave energy increases linearly with time for the wind case, 
and reaches $2 \times 10^{50}$ erg at the deceleration time. We note that solutions with a 
smaller density also fit the data well (and also underestimate the X-ray fluxes) and the
deceleration phase might be delayed for an even longer time.

Radio and X-ray  afterglow observations of GRB 060218  have been
reported in \citet{campanaetal06, soderbergetal06, evansetal07, evansetal09}.
The radio light curve at 8.46 GHz decays as $\propto t^{-0.8}$
\citep{soderbergetal06}. We find that this radio band is between $\nu_a$ and
$\nu_c$ ($\nu_m$ is below $\nu_a$). Using eq. (\ref{F_nu_coasting}), we find that a value
of $p \approx 2.6$ fits the decay of the radio observations, and this
is the reason we present the light curves for this value of $p$ in Fig. \ref{fig3}.
We find an overall good fit to the radio observations (however, some upper
limits at 1.43, 4.86 and 8.46 GHz are consistent only
to within a factor of $\lae 2$).
 We also find 
$\nu_a \sim 4$ GHz at 5 days, and that $\nu_m$ is about ten times smaller than
$\nu_a$, consistent with the spectrum at 5 days presented in \citet{soderbergetal06}. 
With that model the predicted X-ray emission is below the observed 
value by at least a factor of $\sim 10$. 

As in the case of GRB 980425, we lower the value of $p$ to obtain a rough
estimate if a concave electron spectrum may fit the
radio and X-ray data simultaneously. This  increases the predicted X-ray flux, although, it makes
the light curve shallower, see eq. (\ref{F_nu_coasting}), which is inconsistent with
the observed decay $\propto t^{-1.2 \pm 0.1}$
\citep{campanaetal06}. Nevertheless, the discrepancy between
the predicted and observed X-ray flux shrinks with a smaller value of $p$.

Even if such a modification would be capable of explaining the X-ray flux, 
the predicted X-ray spectrum is
$\beta_x = (p-1)/2 \approx 0.8$.
This stands in contrast with the observed X-ray spectrum $\beta_x \approx 2.2 \pm 0.2$
\citep{soderbergetal06}. Such a steep spectrum cannot be explained within this
model, unless the electron spectrum has a cut-off at energies of the X-ray
emitting electrons. This solution is contrived, and we conclude that the X-ray
and the radio are most likely of different origin.

If faster shells than the breakout shell are present, then the blast wave
would start decelerating at much earlier times and it can be relativistic for
long enough for the decreasing radio emission to be consistent also with a
constant density medium. However, in that case $\nu_a< 1.43$ GHz $< \nu_m$ at
5 days, which leads to a flux at 1.43 GHz that is at least a factor of $\sim 3$
brighter than the upper limit at that time. This is marginal even if
there is strong scintillation, but cannot be completely ruled out. The X-ray
emission in that case is still underpredicted by the model if a single
electron power-law is assumed and the difficulty to reproduce the X-ray
spectrum persists.

\subsection{GRB 100316D}

\begin{figure*}
\begin{center}
\includegraphics[width=16cm, angle = 0]{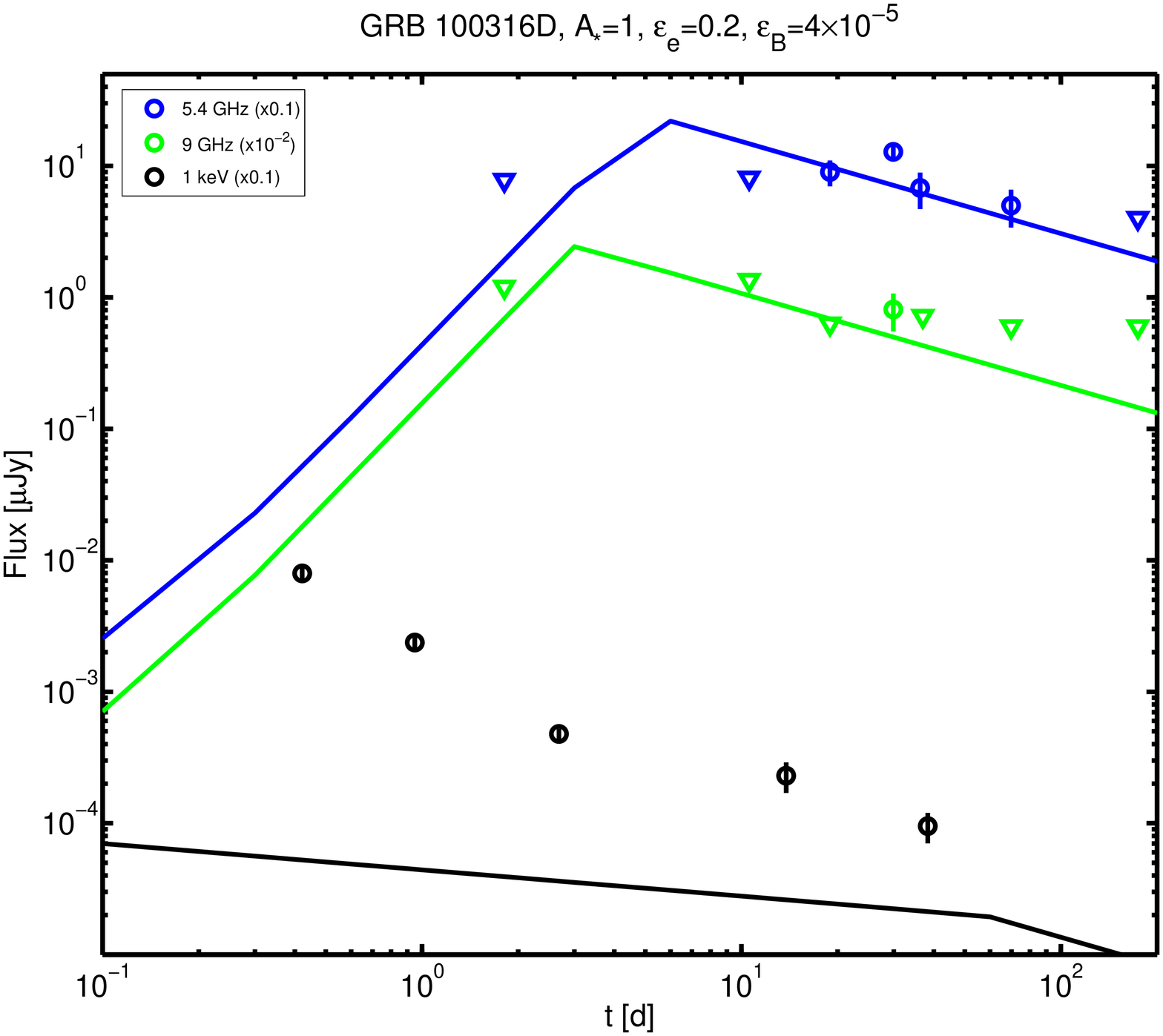}
\end{center}
\caption{The observed and predicted radio and X-ray afterglows of GRB 100316D.
  The fit shown uses $E_{f,0} = 2 \times 10^{50}$ erg, $\gamma_{f,0}=1.3$, 
  $p=2.4$, $\epsilon_e=0.2$ and $\epsilon_B=4 \times 10^{-5}$.
  This model corresponds to the synchrotron emission 
  from the breakout shell during the coasting phase for a wind with $A_*=1$, 
  which can qualitatively fit the radio observations 
  (we note that the radio data are sparse and only available in two
  frequencies, so drawing strong conclusions is difficult).
  However, the model underpredicts the X-ray emission (and, even if the X-ray
  would not be underpredicted, the model cannot explain the steep observed X-ray spectrum).}
\label{fig4} 
\end{figure*}


The last burst in our sample is  GRB 100316D, whose X-ray data are presented in
\citet{evansetal07, evansetal09, starlingetal11}, and additional X-ray data and
the  radio data are presented  in \citet{marguttietal13}.  Since this  burst is
almost identical to  GRB 060218, and it is explained  with the same parameters
within the relativistic shock breakout model, we use the same afterglow model,
namely  synchrotron emission from  a coasting blast  wave in a  wind medium.
The radio data of this burst are sparse, available only in two frequencies
 and  possibly affected  by  scintillation. Therefore, it is not 
 very constraining and many models can fit the
radio data. Here we present one such model in Fig. \ref{fig4}, which
fits the radio data reasonably well. Most importantly, $\nu_a$ is not well 
constrained, e.g., \citet{marguttietal13} suggest that $\nu_a
\sim 5$ GHz at $\sim 30$ days, whereas in the model presented here this value
of $\nu_a$ is obtained at $\sim 6$ days. As a result the radius and thus the 
velocity of the radio emitting region can significantly 
vary between various acceptable models. 

The  model underpredicts X-ray flux compared with  the observed value by a factor
of more than $\sim 4$. As  for GRB 060218, a smaller value of $p$
(concave electron spectrum) increases the predicted X-ray flux, but the
X-ray light curve is too shallow and it is inconsistent with the observed decay 
$\propto t^{-0.87 \pm 0.08}$ \citep{marguttietal13}. 
At the same time the observed X-ray spectrum, $\beta_x \approx 2.49 \pm
0.26$ \citep{marguttietal13}, is similar to that of GRB 060218 and it indicates
that also here the X-ray and the radio are most likely generated by different
processes.

The blast wave moves with a constant velocity and its energy increases linearly
with time until it reaches $2 \times 10^{50}$ erg at the deceleration time 
of $\sim 200 $ d. We note that solutions with a smaller density also fit the 
data well (and also underestimate the X-ray fluxes) and the deceleration phase 
might be delayed for an even longer time.  If faster shells than the breakout 
shell would be present, then an ISM or wind medium would also marginally fit 
the radio data, since the data are not very constraining.

\citet{soderbergetal06, fanetal06} have modeled the radio
emission of GRB 060218, and \citet{marguttietal13} the one of GRB 100316D 
in the context of the external forward shock emission. 
The external densities they find are different than ours. However, here 
the afterglow is produced in the coasting phase,
whereas they considered a decelerating blast wave.

\section{Constraints on the X-ray source} \label{Xray_constraints_section}

The X-ray  afterglows in GRB 060218  and GRB 100316D are  most likely not
generated  by   the  same   process  that   produces  their   radio  emission.
Nevertheless, several  strong constraints on  the X-ray source can  be derived
from simple considerations. These constraints are regardless of whether {\it ll}GRBs
are shock breakouts or not, or even of  the exact models of both the radio and
the X-ray emission. 

In both bursts the total  energy emitted in X-rays by $10^6$ s  
is $\sim 10^{48}$ erg (assuming isotropy;  see below). Thus, the first
general constraint is that  for any reasonable radiative efficiency, the
source energy at that time is $>10^{49}$  erg, and most  likely significantly  higher. The
X-ray energy output is roughly constant for every  logarithmic time interval,
implying that the energy constraint at $10^4$ s is lower by about an order of magnitude.

A second constraint follows from the fact that the gamma-ray emission,
which led to the bursts' detection, is not highly collimated, 
namely opening angle $>30^{\circ}$ and inverse beaming factor $<10$
\citep[e.g.,][]{soderbergetal06}.  Given that  the X-ray  emission is  seen in
both bursts,  it is most  likely not highly  collimated as well.  This implies
that the source cannot be a highly relativistic narrow jet. In fact, the radio
emission implies that  the X-ray source cannot be highly  relativistic even if
it is not narrowly collimated. The  reason is that without being collimated it
cannot run ahead of the radio source, otherwise it will sweep the circum-burst
medium, thereby  preventing the interaction  that leads to the  radio emission.
Now, the radio  source, at least in  GRB 060218 where $\nu_a$  is measured, is
mildly relativistic\footnote {The velocity of the radio source is robust as it
is  independent of  the  exact interaction  details and  depends  only on  the
assumption of  a synchrotron  emission and on  the identification  of $\nu_a$.
\citet{soderbergetal06} measures  $\nu_a \approx 4$ GHz  and $F_{\nu,a} \approx
0.4$ mJy at 5 days for GRB 060218. Using \citet{barniolduranetal13} we find that 
if we require that the energy in the radio emitting region is $\lae 10^{50}$ and we allow for a (large)  
uncertainty in the measurement of $\nu_a$ by a factor of $\sim 2$, then these 
observations imply a Lorentz factor of the radio emitting region
$<2$ at 5 days.}. Therefore, the X-ray source is at most
mildly relativistic, implying  that the X-ray emission radius, $R_x$,  
is limited to $R_x \lae  ct$.

On the other, hand the X-ray emitting radius cannot be too low.  
The line profiles of SN 2006aj (the SN that accompanied GRB 060218) 
do not show signs of anisotropy, and there is no polarization ($<2$ per cent at the 
3-$\sigma$ level) either in the lines nor in the continuum,
implying that the bulk of the SN ejecta has no strong deviation from spherical
symmetry \citep{mazzalietal07}.
This ejecta is about a solar mass moving at a velocity $\gae 0.05$c 
\citep{pianetal06,Bufanoetal12} that lies between the observer and the center of 
the explosion. No radiation that is emitted behind this mass can escape to the observer 
on a time-scale that is shorter than about a week (since the diffusion time is too long). 
Therefore, the X-ray source must be  at all time {\it ahead of} the bulk of the SN 
mass\footnote{In principle, there may be one  line of sight (or more) where there is a `hole' 
in the ejecta (e.g., as may be drilled by a jet), but such a hole, if exists, must be narrow 
enough not to leave an  imprint of anisotropy. It is therefore highly unlikely that the 
X-ray source, which cannot be not narrowly collimated, is seen through such a hole.}.
Thus, together with the limit that the radio
emitting region is ahead of the X-ray source, we obtain $0.05ct  \lae R_x \lae ct$.  Given that the
afterglow is observed over two decades in time, the radius of the region where
X-rays are generated must increase in time together with the SN ejecta. 
This constraint implies that the X-ray source is most likely 
within the explosion ejecta, between the bulk of the SN mass (the slowest part of
the ejecta) and the leading radio source (the fastest part of the ejecta).
This  disfavors emission models that are powered by internal dissipation in an 
unrelated component, such as internal shocks or magnetic reconnection in an 
outflow that is unrelated to the one  producing the SN and the radio emission.

The constraints derived above are non-trivial  and cannot be easily met by most types
of X-ray sources. Specifically, they are most difficult to reconcile with
any source that gains its energy continuously from a central engine,
as has often been suggested \citep[e.g.][]{soderbergetal06,fanetal06,marguttietal13}.
The main problem in such sources  is that  the  energy which  is
generated in the center must penetrate through several solar masses of SN
ejecta, while the source is not relativistic and it is not narrowly collimated
($>30^{\circ}$). Furthermore, this should take place without the energy being dissipated 
below the photosphere and without leaving any  major imprint on the  
asymmetry inferred from the SN line profiles.      

We note that a simple ``inward'' extrapolation in energy (from the energy of the fast 
moving material to the SN energy) described at the end of Section \ref{NR phase} does not work. 
Namely, the energy-velocity relation is too shallow compared to that seen in regular 
SNe and to that expected in a simple spherical blast wave that propagates in a typical 
star (e.g., \citealp{soderbergetal06,marguttietal13}). This is true even when one 
accounts for the shallow relativistic slope described here and in previous works (e.g., 
\citealp{tanetal01}). This probably implies that the explosion is not fully spherical (e.g., 
\citealp{couchetal11, matzneretal13, salbietal14}), or that the stellar structure 
is different than in typical SNe, or a combination of the two. A likely explanation for 
this point is discussed in a future work (Nakar, in preparation). 
Regardless of the reason for the shallow energy profile, the observations 
indicate that the asphericity, if exists, is not high (inverse beaming factor $<10$) 
and that the spherical approximation for the gamma-ray emission and the 
following blast wave is reasonable. Note that if the breakout deviates from 
spherical symmetry this may result in breakout velocities that are slower in 
some parts of the stellar surface \citep{matzneretal13, salbietal14}. But, 
unless the breakout asphericity is high, the breakout from most of the stellar 
surface will not be strongly affected; neither the resulting gamma-rays nor 
the following blast wave.

\section{Discussion and Conclusions} \label{final_section}

We have calculated the synchrotron emission from the
forward shock that arises during the interaction between the stellar ejecta
and the circum-burst medium following a  relativistic shock breakout. 
The hydrodynamics of the shock breakout ejecta is described following  \citet{nakamura06}, 
where we assumed that the progenitor has a standard radiative stellar
envelope,  characterized by a  polytropic index $n=3$. 
For the subsequent blast wave evolution we assume that the
circum-burst  density profile  is either  constant (ISM)  or $\propto  r^{-2}$
(wind). The main difference between this  model and the standard
GRB afterglow model  is that here energy  is continuously injected into  the slowly
decelerating blast wave  by slower ejecta whose profile  is dictated by the
SN shock that unbinds the stellar envelope.
Once the behavior of the blast wave is determined the emission follows 
the standard GRB synchrotron  afterglow model    
\citep{sarietal98}.

We have applied this  model to  the four  {\it ll}GRBs with  good data, 
GRBs 980425, 031203, 060218 and  100316D. The prompt emission 
provides us the initial conditions (the  energy and
Lorentz factor profiles)  of the ejecta \citep{nakarandsari12}. We first  fit, for
each burst, the free parameters of the emission model to 
the observed radio  afterglow and then check if the  same model parameters can
also  explain  the observed  X-ray  afterglow.  We  find  that the  model  can
successfully explain the radio emission of all four {\it ll}GRBs. In some {\it ll}GRBs, the
X-ray emission  can also be  generated by the same  population of electrons
that   generate   the   radio   emission,    while   in   others   it   cannot.

For GRB 031203, we find that the simplest model can successfully fit {\it both}
the observed  radio and X-ray afterglows light curves and spectra with a single set  of parameters, if
the external  medium is  an ISM. This burst is the
only {\it ll}GRB for  which we expect the  blast wave producing the  afterglow to be
relativistic, similarly to regular GRBs, and indeed the parameters we find for
the external density  and microphysical parameters are similar  to those found
in regular GRBs. This is the first  time that {\it both} the prompt and the afterglow
emission can be explained for a single {\it ll}GRB: The prompt emission
is  produced by  the  relativistic shock  breakout signal  and  the radio  and
X-ray afterglow are explained by its accompanying external shock. 

For GRB 980425, a model with a wind medium fits the radio data. The
density of the  wind is such that  the blast wave is already  Newtonian at the
time of the  first observation. Using the simplest model,  the same parameters
that fit the radio  data result in an X-ray spectrum that is  too soft and an
X-ray flux that is too low.  However, the model can simultaneously explain the
radio  and  X-ray  observations  (both the light  curve  and the spectrum)  with  a  minor
modification  that  relaxes the  assumption  of  a single  power-law  electron
distribution. A concave electron distribution with a
logarithmic  derivative of  $p\approx 2.8$  for the radio  emitting electrons  and
$p\approx 2.1$  for the X-ray  emitting electrons can fit both observations. Interestingly such
a   spectrum is  predicted by
some {\it Fermi} acceleration models in  Newtonian shocks, due to nonlinear feedback
of the  accelerated particles on the  shock structure \citep[e.g.,][]{Bell87}.

For GRB 060218  and 100316D, the relativistic shock breakout  model predicts a
mildly relativistic  breakout that  is  consistent with the absence  of ejecta
faster than  the breakout shell. In  that case the breakout  shell decelerates
after a  long time,  and the  observed afterglow  is produced  via synchrotron
emission in the coasting phase. The  observed radio afterglow, which decays at
late times, is consistent only with  a wind medium (in the  ISM case all
light curves rise  during the coasting phase).  This model is able  to fit the
radio data.  However, with  the same parameters,  it underestimates  the X-ray
flux. Moreover, the afterglow X-ray spectra in these two bursts are very steep.
In fact, it is very peculiar compared  to other GRB afterglows. Yet, the light
curve  in  both  cases  (a  very bright  simple  power-law,  roughly  $\propto
t^{-1}$) resembles that of regular GRBs and is uncommon in ordinary SNe. The
steep spectra implies,  first, that it is highly unlikely  that the X-rays and
the radio  are generated by the  same population of electrons,  as was already
found earlier  \citep{fanetal06,soderbergetal06,marguttietal13}.
Secondly, the  combination of the  X-ray spectra and  light curves in  these two
bursts, suggests that  the X-ray source is uncommon to  either regular GRBs or
regular  SNe.

We  stress that  while we  have found  models that  are consistent  with the  radio
emission of all four {\it ll}GRBs, these are by no means the only consistent models.
First, within the  synchrotron forward shock model there is  enough freedom to
allow for a range of parameters that  fit the observations. Secondly, we did not
attempt to calculate  here the synchrotron emission from the  reverse shock or
the contribution  of inverse Compton emission  from either the forward  or the
reverse shocks. It is most likely  that there are solutions where the observed
radio emission  is dominated by one  of these components (e.g.,  reverse shock
synchrotron).

The main puzzle  that remains is the  source of the X-ray  afterglow in {\it ll}GRBs
060218 and  100316D. We draw  here some general  constraints on the  source of
this  emission, finding that it is not  narrowly  collimated and that it is not
ultra-relativistic. The  source itself  is most likely  moving  at velocity
$>0.05$c and  a Lorentz  factor  $<2$, and it  is  located  at all time somewhere within  the
ejecta, between the bulk of the SN mass and the leading radio source.
Previous  authors invoked  the existence  of  a continuous  central engine  to
explain the X-ray afterglow \citep{soderbergetal06,fanetal06,marguttietal13}, 
without  discussing how such a source produces this emission. The constraints we 
derive here show that  long lasting central engine  activity is
unlikely to  be the correct solution.  The major difficulty for  this model is
the need  for the  central engine  energy, that is not  narrowly collimated,
to penetrate  through the  opaque SN
ejecta without leaving a  trace of
anisotropy in the spectral  line profiles \citep{mazzalietal07}. Additionally,
the combination of the X-ray light curve  and spectrum seen in these bursts is
unique and does  not resemble any other  cases where we have a  good reason to
suspect that they are a result of  central engine activity. In 
our view, given that the X-ray source is probably moving together with the  ejecta,
the most  natural scenario is that the X-ray source is within  the
interaction layer that is also responsible for  the radio emission.  
Since the X-ray and  the radio do not arise from the same  process, 
it is possible that either the long lived reverse shock (within a refreshed shocks scenario) 
or inverse Compton emission can explain these unique X-ray signatures.   

\section*{Acknowledgements}

RBD dedicates this work to the memory of Francisco Pla Boada. RBD thanks useful
discussions with Paz Beniamini, Raffaella Margutti, Lara Nava and Rongfeng Shen.
We thank Christopher Matzner for useful comments on the manuscript.
This work was supported by an ERC advanced grant (GRB), by an ERC starting grant (GRB/SN), by the 
I-CORE Program of the PBC and the ISF (grant 1829/12), ISA grant and ISF grants.
This work made use of data supplied by the UK Swift Science Data Centre 
at the University of Leicester.

\end{document}